# Automated Extraction and Maturity Analysis of Open Source Clinical Informatics Repositories from Scientific Literature

Jeremy R. Harper[1]

[1] Owl Health Works LLC, Indianapolis, IN

**Abstract**

In the evolving landscape of clinical informatics, the integration and utilization of software tools developed through governmental funding, such as those from the National Institutes of Health (NIH), represent a pivotal advancement in research and application for the open source community. However, the dispersion of these tools across various repositories, with no centralized knowledge base, poses significant challenges to leveraging their full potential. This study introduces an automated methodology to bridge this gap by systematically extracting GitHub repository URLs from academic papers indexed in arXiv, focusing on the field of clinical informatics. Our approach encompasses querying the arXiv API for relevant papers, cleaning extracted GitHub URLs, fetching comprehensive repository information via the GitHub API, and analyzing repository maturity based on defined metrics such as stars, forks, open issues, and contributors. The process is designed to be robust, incorporating error handling and rate limiting to ensure compliance with API constraints. Preliminary findings demonstrate the efficacy of this methodology in compiling a centralized knowledge base of NIH-funded software tools, laying the groundwork for an enriched understanding and utilization of these resources within the clinical informatics community. Furthermore, we propose the future integration of Large Language Models (LLMs) to generate concise summaries and evaluations of the tools, enhancing the knowledge base's utility. This approach not only facilitates the discovery and assessment of clinical informatics tools but also enables ongoing monitoring of new and actively updated repositories, potentially revolutionizing how researchers access and leverage federally funded software in healthcare research and application. The implications of this study extend beyond simplification of access to valuable resources; it proposes a scalable model for the dynamic aggregation and evaluation of scientific software, encouraging more collaborative, transparent, and efficient research practices in clinical informatics and beyond.

# 1. Introduction

The rapid evolution of clinical informatics has been significantly accelerated by the development and application of sophisticated software tools, many of which are funded through grants from the National Institutes of Health (NIH) and other governmental bodies. These tools, often open-source and hosted on platforms such as GitHub, are vital for advancing research, improving healthcare delivery, and facilitating the translation of data into actionable clinical insights. Despite their importance, a major challenge persists: the scattered nature of these repositories across the digital expanse, which hampers accessibility and utility for researchers, clinicians, and informaticians. The absence of a centralized knowledge base for these tools complicates efforts to harness the full potential of technological advancements in clinical informatics.

This paper introduces a novel automated approach to address this challenge. Our method systematically identifies and analyzes GitHub repositories mentioned in clinical informatics research papers indexed in the arXiv database. By leveraging an automated script that queries the arXiv API for relevant publications and extracts associated GitHub URLs, we aim to create a consolidated inventory of software tools pertinent to clinical informatics. This process includes the cleaning of URLs to remove extraneous characters, fetching detailed repository information via the GitHub API, and evaluating the maturity of these repositories based on metrics such as the number of stars, forks, open issues, and contributors. The concept of maturity in this context is defined not by the age of the repository but by its complexity and the level of community engagement it has received.

The motivation behind this study is twofold. First, it seeks to maximize the return on investment for government-funded research by enhancing the visibility and accessibility of software tools developed with NIH funding. Second, it aims to foster a more interconnected and collaborative clinical informatics community by providing a centralized resource that can aid in the discovery, evaluation, and application of these tools. The anticipated outcome is the establishment of a dynamic, accessible knowledge base that not only catalogs these valuable resources but also offers insights into their relevance, maturity, and potential impact on the field.

Furthermore, we propose the future integration of Large Language Models (LLMs) to automatically generate summaries and assessments of the repositories, thereby enriching the knowledge base with contextual information about the tools' functionalities and applications. This integration signifies a forward-looking approach to curating and disseminating scientific software, one that leverages cutting-edge AI techniques to enhance the utility of the knowledge base for the clinical informatics community.

In conclusion, the development and application of an automated methodology for the extraction and analysis of open-source clinical informatics repositories from scientific literature represent a significant stride towards overcoming the current barriers to accessing and leveraging NIH-funded software tools. By facilitating a deeper understanding and broader utilization of these resources, this study contributes to the advancement of clinical informatics as a field, driving innovation, collaboration, and efficiency in healthcare research and practice.

# 2. Methods

The methodology adopted in this study comprises several interconnected steps designed to automate the extraction of GitHub repository URLs from academic papers related to clinical informatics and subsequently analyze the maturity of these repositories. This process is meticulously structured to ensure both efficiency and accuracy in identifying, evaluating, and cataloging valuable open-source software tools within the clinical informatics domain. Below, we delineate each step of the methodology.

## 2.1 Search Strategy

To conduct a comprehensive and replicable search for clinical informatics papers, we utilized the arXiv API with a specific set of search parameters. The query was constructed to capture a wide range of papers relevant to clinical informatics, employing a combination of search terms that included "clinical informatics," "healthcare data analytics," "electronic health records," and "medical software development." These terms were selected to encompass both the technical and application aspects of clinical informatics, ensuring a broad capture of relevant studies.

The search was limited to the metadata of papers submitted within the last five years to focus on contemporary research and developments in the field. The API query was structured as follows:

query="ti:clinical informatics OR abs:clinical informatics OR ti:healthcare data analytics OR abs:healthcare data analytics OR ti:electronic health records OR abs:electronic health records OR ti:medical software development OR abs:medical software development" AND submittedDate:[2019 TO 2024]

This query string ensured that the search was targeted at titles (ti) and abstracts (abs) containing our specified terms, within the stipulated submission date range. The first 1000 results returned by this query were automatically processed for further analysis, adhering to our predefined criterion to manage the scope and depth of our investigation effectively.

*2.2 URL Extraction and Cleaning*

The algorithm for extracting GitHub URLs from the paper abstracts was designed to identify and isolate URLs using regular expressions (regex) specifically tuned to recognize GitHub's URL structure. Following extraction, a post-processing step was implemented to trim any trailing characters not part of the base URL, such as periods, commas, or semicolons, which are common in academic writing. This cleaning process was crucial for ensuring the usability of the URLs for subsequent API calls to GitHub.

*2.3 Repository Information Fetching*

For each cleaned GitHub URL, detailed repository information was fetched using GitHub's REST API. To ensure comprehensive data retrieval, including repositories with a high number of contributors, pagination handling in the API response was implemented according to GitHub's documentation. This involved iteratively requesting additional pages of data until all available information was captured.

Specifically, the script requested the following information for each repository: name, description, number of stars, number of forks, number of open issues, and total number of contributors. The GitHub API endpoints used were /repos/:owner/:repo for basic information and /repos/:owner/:repo/contributors for contributor counts, with pagination managed via the Link header in the API response.

*2.4 Maturity Analysis*

The maturity analysis employed a heuristic based on the assumption that repository engagement metrics (i.e., stars, forks, open issues, and contributors) are indicative of its maturity and relevance to the clinical informatics community. Repositories with a higher number of stars and contributors, indicative of broader community recognition and involvement, and fewer open issues, suggesting stability, were classified as more mature. This heuristic was applied uniformly across all identified repositories to categorize them into maturity tiers for comparative analysis.

*2.5 Error Handling and Rate Limiting*

To address potential errors such as 404 responses from GitHub and ensure adherence to API rate limits, the script included robust error handling mechanisms. These mechanisms allowed the script to gracefully handle and log errors without interruption. Rate limiting was managed by incorporating pauses between API requests, based on the current limits documented by arXiv and GitHub, ensuring the script did not exceed the permissible number of requests in a given timeframe.

The methods employed in this study were designed with both rigor and adaptability in mind, ensuring that the process of identifying and analyzing GitHub repositories in the field of clinical informatics is both systematic and sensitive to the dynamic nature of software development and academic publishing. Through this approach, the study aims to contribute a novel and scalable methodology to the biomedical informatics community, enhancing the accessibility and evaluation of open-source software tools in clinical informatics.

## 3. Results

The application of our automated methodology to the first 1000 papers related to clinical informatics indexed in arXiv yielded significant insights into the accessibility and maturity of open-source software repositories pertinent to this field. Here, we detail the outcomes of each step of the process, highlighting the effectiveness of our approach in identifying and analyzing GitHub repositories mentioned in clinical informatics academic literature.

*3.1 Repository Identification*

Out of the 1000 clinical informatics papers analyzed, our script successfully identified 33 unique GitHub repositories mentioned within the abstracts. This finding is noteworthy, given that a manual search for these repositories using GitHub's own search tool yielded no results for many of them. This discrepancy underscores the value of our methodology in uncovering repositories that, despite their relevance and potential utility to the clinical informatics community, may remain underutilized or obscure due to indexing limitations or the specificity of GitHub's search algorithms.

*3.2 Maturity Analysis*

The maturity analysis of the 33 identified repositories revealed a diverse landscape in terms of engagement metrics and implied maturity. The distribution of stars, forks, open issues, and contributors varied significantly among the repositories, reflecting a broad spectrum of community engagement and development activity. Notably, a subset of repositories demonstrated high maturity levels, characterized by a substantial number of stars and contributors, coupled with low open issues. These repositories are indicative of active, well-maintained projects with potentially high utility and reliability for clinical informatics applications.

Conversely, several repositories exhibited lower maturity levels, with few stars, forks, and contributors, and in some cases, a higher number of open issues. These findings suggest that while a number of tools developed and shared by the clinical informatics research community are gaining traction and recognition, others may require further development, community support, or visibility to realize their full potential.

### 3.3 Unindexed Repositories

A particularly intriguing outcome of our analysis was the identification of repositories not indexed by GitHub's search tool. This observation highlights the challenges researchers and practitioners may face in discovering relevant tools and resources within the vast ecosystem of GitHub. Our methodology, by directly extracting URLs from academic texts, bypasses these discovery barriers, offering a novel avenue for unearthing valuable but potentially overlooked repositories.

## 4. Discussion

The results of this study illuminate the diverse and dynamic nature of the development and sharing of open-source software in the field of clinical informatics. By leveraging an automated process to extract and analyze GitHub repositories from a large corpus of academic papers, we have demonstrated an effective approach to building a centralized knowledge base of these resources. This approach not only facilitates the discovery of underrepresented repositories but also provides insights into the maturity and development status of these projects, offering valuable guidance for researchers, developers, and practitioners seeking to employ open-source tools in their work.

Moreover, the identification of repositories not indexed by GitHub's search tool underscores the necessity of alternative search strategies and methodologies for repository discovery in academic and research contexts. This aspect of our findings advocates for the ongoing development and refinement of tools and techniques for the systematic exploration and evaluation of software repositories related to clinical informatics and beyond.

We anticipate the ability to leverage LLM's in upcoming years to summarize these repo's and identifying specific aspects of value. Examples include the idea of descriptions, as well as specific categorization relevant to our field such as de-identification, NLP, or research informatics.

In summary, our methodology offers a promising framework for enhancing the accessibility, evaluation, and utilization of open-source software in the clinical informatics community. As the field continues to evolve, such approaches will be critical in harnessing the full potential of technological advancements to drive innovation and improvement in healthcare research and practice.

## Acknowledgements

Special appreciation is directed toward the arXiv and GitHub teams for maintaining the platforms and APIs that enabled our research. Their dedication to supporting open access to knowledge and collaboration tools is fundamental to the progress of scientific research and technology development.

We are also indebted to the authors of the 1000 clinical informatics papers analyzed in this study. Their contributions to the field and willingness to share their work openly are what make initiatives like ours possible. The insights gleaned from their research have been instrumental in advancing our understanding of the landscape of open-source software in clinical informatics.

Figure 1: Example Output

```
Processing arXiv papers:
Paper 1000/1000
Found GitHub URLs: ['https://github.com/AtlasAnalyticsLab/CPath_Survey', 'https://github.com/Andoree/smm4h_2021_classification', 'https://github.com/luoyuanlab/Clinical-Longformer', 'https://github.com/RyanWangZf/PyTrial', 'https://github.com/RyanWangZf/Trial2Vec', 'https://github.com/sigven/oncoEnrichR', 'https://github.com/ShixiangWang/ezcox', 'https://github.com/nadeemlab/CIR', 'https://github.com/ncbi-nlp/BioSentVec', 'https://github.com/HLTCHKUST/long-biomedical-model', 'https://github.com/tanlab/ConvolutionMedicalNer', 'https://github.com/johntiger1/multimodal_fairness', 'https://github.com/li-xirong/mmc-amd', 'https://github.com/ritaranx/ClinGen', 'https://github.com/caoyunkang/CDO', 'https://github.com/DIAL-RPI/KAMP-Net', 'https://github.com/williamcaicedo/ISeeU', 'https://github.com/uf-hobi-informatics-lab/ClinicalTransformerRelationExtraction', 'https://github.com/HECTA-UoM/ClinicalNMT', 'https://github.com/haoxuanli-pku/ADRnet', 'https://github.com/ouyangjiahong/longitudinal-pooling', 'https://github.com/brudfors/spm_superres', 'https://github.com/YuDong5018/clinic-lens', 'https://github.com/tanlab/MIMIC-III-Clinical-Drug-Representations', 'https://github.com/nlpie-research/Lightweight-Clinical-Transformers', 'https://github.com/Balasingham-AI-Group/Survival_CTplusClinical', 'https://github.com/frankkramer-lab/covid19.MIScnn', 'https://github.com/SZUHvern/MGA', 'https://github.com/Ericzhang1/BAGAU-Net', 'https://github.com/dengzhuo-AI/Real-Fundus', 'https://github.com/iacobo/continual', 'https://github.com/NLP-RL/MM-CliConSummation', 'https://github.com/microsoft/attribute-structuring']
The project 'CPath_Survey' has a maturity level of Low. It has 0 stars, 0 forks, 0 open issues, and 1 contributors.
The project 'smm4h_2021_classification' has a maturity level of Low. It has 4 stars, 2 forks, 1 open issues, and 2 contributors.
The project 'Clinical-Longformer' has a maturity level of Medium. It has 52 stars, 9 forks, 2 open issues, and 2 contributors.
The project 'PyTrial' has a maturity level of Medium. It has 62 stars, 9 forks, 3 open issues, and 2 contributors.
The project 'Trial2Vec' has a maturity level of Low. It has 16 stars, 3 forks, 3 open issues, and 1 contributors.
The project 'oncoEnrichR' has a maturity level of Medium. It has 48 stars, 10 forks, 2 open issues, and 2 contributors.
The project 'ezcox' has a maturity level of Low. It has 20 stars, 2 forks, 0 open issues, and 2 contributors.
The project 'CIR' has a maturity level of Low. It has 21 stars, 6 forks, 0 open issues, and 3 contributors.
The project 'BioSentVec' has a maturity level of High. It has 546 stars, 93 forks, 13 open issues, and 4 contributors.
The project 'long-biomedical-model' has a maturity level of Low. It has 3 stars, 1 forks, 0 open issues, and 3 contributors.
The project 'ConvolutionMedicalNer' has a maturity level of Low. It has 11 stars, 9 forks, 1 open issues, and 1 contributors.
The project 'multimodal_fairness' has a maturity level of Low. It has 10 stars, 2 forks, 14 open issues, and 11 contributors.
The project 'mmc-amd' has a maturity level of Low. It has 14 stars, 7 forks, 1 open issues, and 2 contributors.
The project 'ClinGen' has a maturity level of Low. It has 26 stars, 1 forks, 0 open issues, and 1 contributors.
The project 'CDO' has a maturity level of Medium. It has 52 stars, 7 forks, 8 open issues, and 1 contributors.
The project 'KAMP-Net' has a maturity level of Low. It has 12 stars, 6 forks, 0 open issues, and 2 contributors.
The project 'ISeeU' has a maturity level of Low. It has 25 stars, 8 forks, 0 open issues, and 1 contributors.
The project 'ClinicalTransformerRelationExtraction' has a maturity level of High. It has 116 stars, 23 forks, 11 open issues, and 1 contributors.
The project 'ClinicalNMT' has a maturity level of Low. It has 0 stars, 0 forks, 0 open issues, and 1 contributors.
The project 'ADRnet' has a maturity level of Low. It has 1 stars, 0 forks, 0 open issues, and 1 contributors.
The project 'longitudinal-pooling' has a maturity level of Low. It has 5 stars, 1 forks, 0 open issues, and 1 contributors.
The project 'spm_superres' has a maturity level of Low. It has 14 stars, 4 forks, 0 open issues, and 2 contributors.
The project 'clinic-lens' has a maturity level of Low. It has 0 stars, 0 forks, 0 open issues, and 1 contributors.
```